\documentclass[aps,twocolumn,preprintnumbers,amsmath,amssymb]{revtex4}

\usepackage{amsmath,amssymb}
\usepackage{bm}
\usepackage{graphicx,subfigure,xcolor}
\usepackage{xcolor}
\usepackage{microtype}
\usepackage{multirow}
\usepackage{booktabs}
\usepackage{fancyhdr}
\usepackage{braket}
\usepackage{mathtools}
\usepackage{comment}
\usepackage{latexsym}
\usepackage{dcolumn}
\usepackage{epsf}
\usepackage{float}
\usepackage{hyperref}

\makeatletter

\renewenvironment{widetext@grid}{%
  \par\ignorespaces
  \setbox\widetext@top\vbox{%
   \vskip15\p@
   \hb@xt@\hsize{%
    \leaders\hrule\hfil
    \vrule\@height6\p@
   }%
   \vskip6\p@
  }%
  \setbox\widetext@bot\hb@xt@\hsize{%
    \vrule\@depth6\p@
    \leaders\hrule\hfil
  }%
  \onecolumngrid
  \let\set@footnotewidth\set@footnotewidth@ii
}{%
  \par
  \twocolumngrid\global\@ignoretrue
  \@endpetrue
}%

\makeatother

\newcommand{\ad}{a^{\dag}}
\newcommand{\bd}{b^{\dag}}
\newcommand{\Tr}{\hbox{Tr}}

\newcommand{\ba}{\begin{array}}
\newcommand{\ea}{\end{array}}
\newcommand{\bqa}{\begin{eqnarray}}
\newcommand{\eqa}{\end{eqnarray}}
\newcommand{\um}{{\bf 1}}

\begin{document}

\title{Deterministic preparation of highly non-classical macroscopic quantum states}

\author{Ludovico Latmiral}
\email{ludovico.latmiral@hotmail.it}
\author{Florian Mintert}

\affiliation{QOLS, Blackett Laboratory, Imperial College London, London SW7 2AZ, United Kingdom}

\begin{abstract}
\bf{We present a scheme to deterministically prepare non-classical quantum states of a massive mirror including highly non-Gaussian states exhibiting sizeable negativity of the Wigner function.
This is achieved by exploiting the non-linear light-matter interaction in an optomechanical cavity by driving the system with optimally designed frequency patterns. Our scheme reveals to be resilient against mechanical and optical damping, as well as mechanical thermal noise and imperfections in the driving scheme. Our proposal thus opens a promising route for table-top experiments to explore and exploit macroscopic quantum phenomena.}
\end{abstract}

\date{\today}

\maketitle

\section*{Introduction}

Non-classicality of mechanical motion has recently been a topic of great interest both theoretically and experimentally as it represents a test ground to address many important questions ranging from quantum-to-classical transition and collapse models \cite{penrose2003, romeroisart2011, bahrami2014} to the interface between quantum mechanics and gravity \cite{pikovski2012, bawaj2015}. While we have extensive literature that has focused on the quantumness of microscopic objects, it is a challenge to deterministically isolate genuine quantum features that can be accessed in experiments, and few experiments with coherent superpositions of quantum objects with large mass exist \cite{hackermuller2003, hackermuller2004}.

Massive mechanical oscillators have been intensively investigated in quantum optomechanics \cite{kippenberg2008, aspelmeyer2014}, and optomechanical cavities are regarded as an optimal framework to make clear comparisons between the predictions of classical theory and their quantum counterparts \cite{mancini1997, bose1997, bose1999, yang2013, latmiral2016a, armata2016}. 
Indeed, they were proven to exhibit a large degree of macroscopicity, $\mu$, defined in terms of the robustness of a coherent superposition against decoherence \cite{nimmrichter2013}.
Optomechanical experiments have reached $\mu=19$ on a scale where the Mach-Zender interference of Cs \cite{isenhower2009} and the Schr\"odinger gedanken experiment are attributed values of $\mu=10.6$ and $\mu\sim 55$ respectively \cite{nimmrichter2013}.

Thanks to their peculiar properties, these systems have been historically studied in the context of force sensing \cite{caves1980, braginsky1995} and for the preparation of non-classical states of the mechanical motion, such as squeezed states \cite{kronwald2013, wollman2015, pirkkalainen2015, lecocq2015, lei2016}, single phonon excitations \cite{rips2012, qian2012, borkje2014} or even Schr\"odinger cat states \cite{bose1997}.
Given the necessary interaction between optical and mechanical degrees of freedom, most control schemes result in the preparation of correlated states.
The reduced state of the mechanical components is then strongly mixed, and a pure (or less strongly mixed) state can be obtained in terms of a measurement on the optical field \cite{abdi2015, liao2016}.
Since such a measurement has random outcomes, such a state preparation is intrinsically probabilistic.
To the best of our knowledge, the only currently existing deterministic protocols rely on equilibration to a stationary state, being based on dissipative state preparation with the potential to prepare superpositions of two wave packets \cite{asjad2014,abdi2016}.

In this paper we consider the deterministic preparation of highly non-classical, motional states via coherent control.
Such a deterministic protocol, that permits to prepare non-stationary states, first of all helps to avoid the additional element of a measurement which is likely to be affected by limited detection efficiencies and dark counts.
Since targeting states with increasing macroscopicity typically implies lower success rates of probabilistic protocols, this shall be helpful, in particular, for the experimental realisation of non-classical states of macroscopic character.
Explicitly, we show how the non-linear light-matter interaction between an electromagnetic field and a movable mirror in an optomechanical cavity can be exploited to deterministically prepare on demand quantum states of the mirror such as squeezed states and non-Gaussian coherent superpositions exhibiting sizeable negativity of the Wigner function. Our control scheme proves to be resilient to several experimental imperfections, permitting maximally non-classical states to be achieved, which makes it ideal for accurate tests of decoherence models and of potential limitations on coherent superpositions of massive objects.

\section*{Results}

We consider an optomechanical cantilever modelled as harmonic oscillator of mass $m$, interacting with a light field through radiation pressure in the single mode approximation.
This provides an accurate description for current experiments \cite{qian2012,wollman2015,aspelmeyer2014}, though the techniques derived in the following also apply to optomechanical systems that are not based on cantilevers, or also more complex models including more degrees of freedom.
The free evolution of the system is given by $\omega_c \ad a + \omega_m \bd b$, where $\omega_m$ ($\omega_c$) is the mechanical (cavity resonance) frequency and $b$ and $\bd$ ($a$ and $\ad$) are respectively the annihilation and creation operator of the mirror (cavity field).
The interaction couples the intensity of the light field with the position of the mechanical element and is described by $H_{int}=-g\ad a(b+b^\dagger)$ \cite{law1995}, where $g=\omega_c\frac{L}{L_c}=k\omega_m$ is the coupling constant,  $L=\sqrt{\hbar/(2m\omega_m)}$ the oscillator length scale, $L_c$ the cavity length at equilibrium and $k=g/\omega_m$ the rescaled coupling.\\
Adding external driving $\xi(t)$ of the cavity, the complete Hamiltonian of the system reads $H=H_0+H_{int}$, with $H_0=\omega_c \ad a + \omega_m \bd b+i\left(\xi(t) \ad -\xi^\ast(t) a\right)$.
Generally the dynamics induces correlations between both subsystems.
A correlated state, however, implies that a mixed quantum state needs to be attributed to each subsystem alone, or that the measurement on one of the subsystems results in the probabilistic preparation of the other.

The goal of the present paper lies in finding driving patterns $\xi(t)$ such that the cubic optomechanical interaction creates non-trivial states of the mirror without cavity-mirror correlations.
In particular, the chosen driving profiles will also ensure that the cavity ends up in its initial state, which will significantly ease the readout subsequent to the state preparation.
Indeed, most of the current state reconstruction techniques of mechanical motional states are achieved through homodyne tomography of a probe light field, \textit{i.e.} the so called \textit{back-action-evading interaction} \cite{zhang2003, lei2016, clark2017}.
It is therefore an essential requirement that the cavity is in its well defined initial state when the read out of the mechanics is performed.

In the limit of weak coupling $k\ll 1$, which is in agreement with state-of-the-art experiments operating at $k\lesssim 10^{-2}$ \cite{kippenberg2008,aspelmeyer2014}, we can solve the dynamics in a perturbative expansion in powers of $k$.
To this end, it is helpful to first find the time-evolution operator $U_0(t)$ induced by the non-interacting time-dependent Hamiltonian $H_0(t)$. 
Since $H_0(t)$ is harmonic, $U_0(t)$ is constructed exactly and it is subsequently used to extract the interaction Hamiltonian in the frame defined by the harmonic motion as $H_I(t)=U_0^\dag(t) H_{int} U_0(t)$, which explicitly reads
\begin{equation}
\label{hint}
\begin{split}
H_I(t)=&-g\Bigl( n_c-(f a^\dag +f^\ast a)+|f|^2\Bigr)X_m(t)\ ,\\
\end{split}
\end{equation}
with $X_m(t)=b^\dag e^{i\omega_m t}+b e^{-i\omega_m t}$, $f=\int_{0}^t dt_1 \xi(t_1) e^{i\omega_c(t_1)}$ and $n_c=a^\dag a$ the number operator of the cavity field.\\
Because of the cubic nature and the time-dependence, it is not possible to analytically solve the generator $V(t,t_0)$ induced by $H_I(t)$, but it can be obtained in the perturbative Magnus series \cite{blanes2009} $V(t,t_0)=\exp\Bigl(-i\sum_j{\cal M}_j(t,t_0)\Bigr)$, where ${\cal M}_1(t,t_0)=\int_{t_0}^t dt_1 H_I(t_1)$, ${\cal M}_2(t,t_0)=-\frac{i}{2} \int_{t_0}^tdt_1  \left[H_I(t_1),{\cal M}_1(t_1,t_0)\right]$ and higher order terms ${\cal M}_j$
satisfy the proportionality ${\cal M}_j(t,t_0)\sim k^j$.

Given the explicit form of $H_I(t)$ in Eq.~\eqref{hint}, the lowest order term ${\cal M}_1$ is an interaction that induces correlations between cavity and mirror.
The higher order expansions ${\cal M}_j$ ($j>1$) will generally also contain both interaction and single-particle terms of mirror or cavity alone.
Since the central goal of our work is deterministic state preparation, we will require that ${\cal M}_1(t)$ and undesired terms in ${\cal M}_j(t)$ ($j>1$) vanish at the final instance in time $t=NT$, after $N$ periods $T=2\pi/\omega_m$ of the mechanical motion.
We will design driving profiles $\xi(t)$ such that all interaction terms and all operators acting on the cavity vanish at $t=NT$, but such that the single-particle terms acting solely on the mirror induce highly non-classical states.

Since for a general time dependent driving $\xi(t)$ it might be difficult to directly integrate the dynamics over $N$ periods, it will prove useful to express the propagator $V(TN,0)$ as
\begin{equation}
\begin{split}
V(TN,0)&=\prod_{s=1}^NV(Ts,T(s-1))=\prod_{s=1}^N\exp(-i{\cal M}^{(s)})\nonumber,
\end{split}
\end{equation}
where it is implied that terms are ordered with decreasing value of $s$ in the product; the ${\cal M}^{(s)}$ are defined via the relation $\exp(-i{\cal M}^{(s)})=V(Ts,T(s-1))$, and can be expanded in the Magnus series ${\cal M}^{(s)}=\sum_j{\cal M}_j^{(s)}$.
Conversely, using Baker-Campbell-Hausdorff relation we can rearrange all terms at the same order in the coupling, \textit{i.e.} ${\cal M}_1(NT,0)=\sum_{s=1}^N{\cal M}_1^{(s)}$ and similarly at higher orders. While there is no reason to expect light-matter correlations and cavity excitation terms to add up to zero at each order $j$ in ${\cal M}_j(NT,0)$, we propose time dependent driving profiles $\xi_s(t)$ resulting in different interaction Hamiltonians $H_I^{(s)}(t)$ in each interval.
With the specific choice $H_I^{(s)}(t)=W_s^\dagger H_I^{(1)}(t)W_s$ (with $W_1=\um$) one obtains $V(TN,0)=\prod_{s=1}^N W_s^\dagger V(T,0)W_s=\prod_{s=1}^N\exp(-i{\cal M}^{(s)})$ with ${\cal M}^{(s)}=W_s^\dagger {\cal M}^{(1)}W_s$.
Since all terms now depend on the $W_s$, which can be chosen freely, we will benefit from this freedom to ensure that any undesired term in ${\cal M}_j$ vanishes or is modified as desired.
As we will see in the following there are clear physically motivated choices for the $W_s$ that achieve the aim, and that translate into rather simple driving profiles.

Due to the large separation of the resonance frequencies of cavity and mirror ($\omega_c/\omega_m \sim O(10^7)$), it is essential to drive the former close to the sidebands with frequencies $\omega_c\pm\omega_m$ to enable the exchange of excitations between the two subsystems.
We will hereafter find suitable profiles such that the mirror evolves into a strongly squeezed state as well as a state with pronounced non-Gaussian and non-classical features.
Apart from an interest in its own, the discussion on strongly squeezed states shall help to exemplify the framework developed above, with simpler algebra than found in the preparation of non-classical states.

Mechanical squeezing is obtained via a bi-chromatic driving with detunings $\pm\omega_m$ with respect to the cavity resonance.
The related driving profile $\xi_\omega(t)={\cal E}e^{-i\omega_c t}(e^{i\omega_m t}+e^{-i\omega_m t})$ with amplitude $\mathcal{E}$ results in the lowest order contribution to the Magnus expansion after one period
\begin{equation}\label{eq:M11}
{\cal M}_{1}^{(1)}=-2\pi k\eta X_cP_m\ ,
\end{equation}
with the dimensionless amplitude $\eta=\mathcal{E}/\omega_m$.
This suggests the particularly simple choice $W_s=\exp(-in_c\varphi_s)$, that rotates cavity operators in phase space by an angle $\varphi_s$.
The corresponding required driving profiles
\begin{equation}\label{drivingprof}
\xi_s(t)={\mathcal E} e^{i\varphi_s} e^{-i\omega_c t}(e^{i \omega_m t}+e^{-i \omega_m t})\ ,
\end{equation}
are obtained by reverse-engineering the derivation of the interaction Hamiltonian (see Eq.~\eqref{hint}) and are rather elementary to implement \cite{thom2013, supplementary}.
In fact, different driving periods differ from each other merely by the phase shift $\varphi_s$, such that
eqs.~\eqref{eq:M11} and \eqref{drivingprof} result in
\begin{equation}
\sum_{s=1}^{N}{\cal M}_{1}^{(s)}=-2\pi\ k\eta\left( \sum_{s=1}^{N}X_c\cos\varphi_s+P_c\sin\varphi_s\right) P_m\ .
\nonumber
\end{equation}
Hence, undesired interaction terms in ${\cal M}_1$ cancel for any choice satisfying $\sum_s\exp(i\varphi_s)=0$.\\
The second order contribution reads ${\cal M}_2(NT,0)=\sum_{s=1}^N{\cal M}_2^{(s)}-\frac{i}{2}\sum_{s>l=1}^N[{\cal M}_1^{(s)},{\cal M}_1^{(l)}]$ and contains correlations and single particle excitation terms of the cavity that vanish upon the condition $\sum_se^{i2\varphi_s}=0$, which eventually motivates the selection $\varphi_s=2\pi (s-1)/N$ (assuming $N>2$).\\
The most important term in ${\cal M}_2$ for the creation of a mechanical squeezed state originates from the commutator $[{\cal M}_1^{(s)},{\cal M}_1^{(l)}]$ and is proportional to $\propto (k\eta)^2\sin(\varphi_s-\varphi_l)P_m^2$.
With the choice $\varphi_s=2\pi (s-1)/N$, the sum over all possible combinations $s>l=1$ reads $\sum_{l<s} \sin\left(\varphi_s-\varphi_l\right) = \frac{N}{2}\cot\left(\frac{\pi}{N}\right)$, which scales $\sim N^2$ and thus becomes sizeable already after few periods of driving.

\begin{figure}[b!]
\centering
\includegraphics[width=0.48\textwidth]{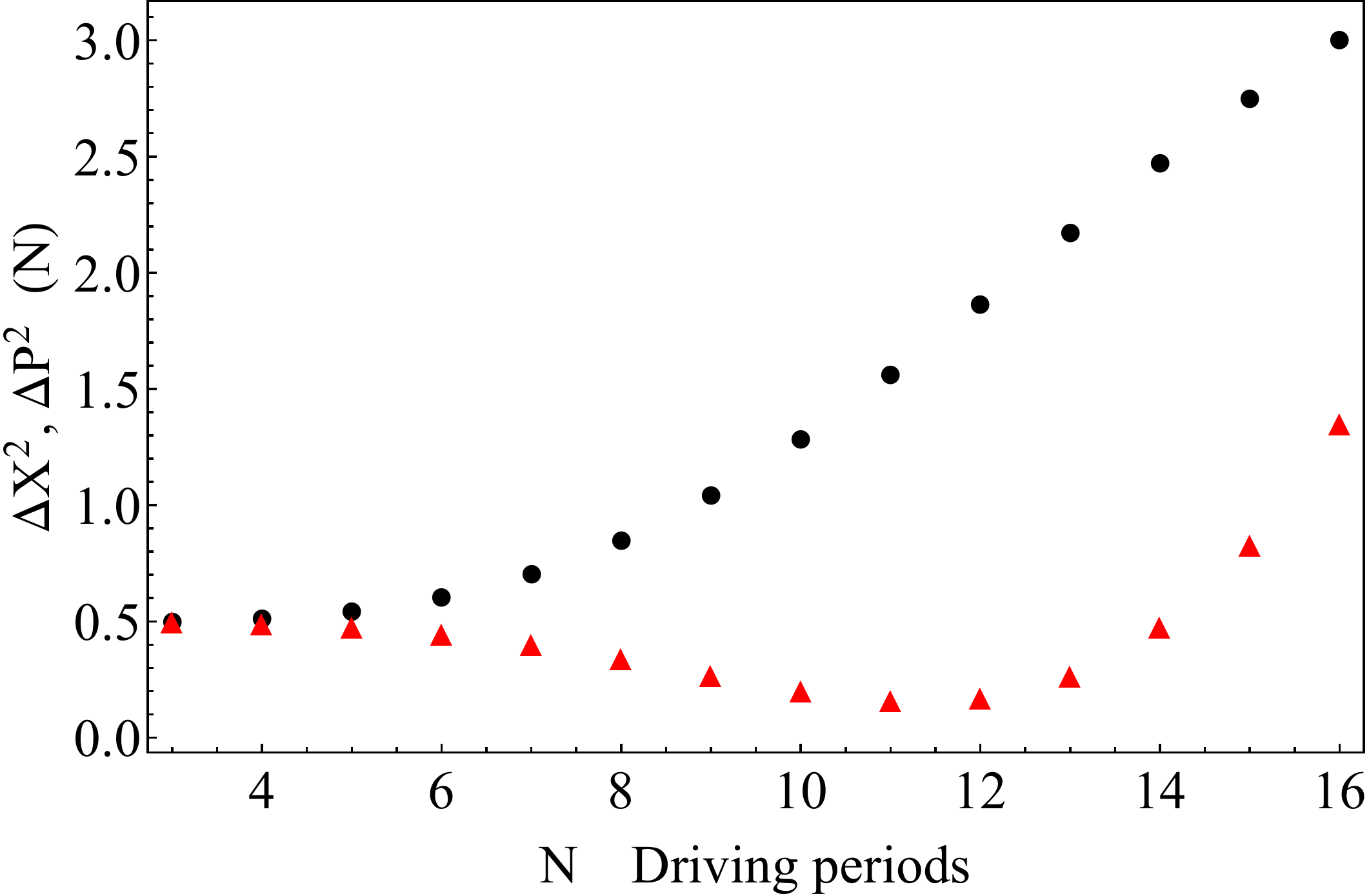}
\caption{Expected values for the quadratures of the mirror as a function of the total driving time expressed in terms of driving periods. Black circles represent $\Delta P^2$ and red triangles $\Delta X^2$, which is squeezed by the evolution operator up to $\Delta X^2=0.16$. Experimental parameters are set as: $\eta=10$, $k=1/400$.}\label{squeezed}
\end{figure}

All-together, we have thus arrived at dynamics, such that no results of an interaction appear at the final instance in time and such that no excitations in the cavity have been created. Up to a global phase factor, which we will henceforth always neglect, the full propagator reads $V_W(TN,0)=V_c(N)\otimes V_m^{(2)}(N)$ with
\begin{equation}
\label{squeezedunitary}
\begin{split}
V_c(N)&=\exp \left( 2\pi i\ N k^2\ ({n}_c^2+7\eta^2 n_c)\right)\ ,\hspace{.3cm}\mbox{and}\\
V_m^{(2)}(N)&=\exp\left(2i \left(\pi k\eta\right)^2 N\cot\left(\frac{\pi}{N}\right) P_m^2\right)\ .
\end{split}
\end{equation}
$V_m^{(2)}(N)$ acts on the mirror only, and can be recast in the form
\begin{equation}
V_m^{(2)}(N)=e^{i\delta\bd b}e^{\frac{1}{2}(\zeta^\ast b^2-\zeta{\bd}^2)}\ ,
\end{equation}
corresponding to a vacuum squeezing operation with parameter
\begin{equation}
\zeta=i\left(2\pi k\eta \right)^2N\cot\left(\frac{\pi}{N}\right)e^{i\delta}\ ,\nonumber
\end{equation}
and followed by a rotation with angle
\begin{equation}
\delta=\arctan\left(\left(2\pi k\eta \right)^2N\cot\left(\frac{\pi}{N}\right)\right)\ . \nonumber
\end{equation}
The quadratic scaling with time (\textit{i.e.} $|\zeta|\sim N^2$) allows substantial squeezing already after a few intervals.
Besides, we should keep in mind that the perturbative regime requires reasonably short propagation times, {\it i.e.} small values of $N$, and the present analysis is valid in the limit $k\ll 1$, as the neglected third order term scales as ${\cal M}_3\sim k^3\eta^2N$.
For a relatively weak interaction, $k=1/400$, and sufficiently strong driving, $\eta=10$, one achieves a squeezing of the position quadrature resulting, after $N=11$ periods, in $\Delta P_m^2 = 1.57$ and $\Delta X_m^2\simeq 0.16$ (see Fig.\ref{squeezed}).

Let us now discuss the creation of non-Gaussian states, which requires to suppress not only interaction effects, but also Gaussian contributions to the dynamics, since these will tend to over-shadow non-Gaussian features.
We will therefore double the detuning as compared to Eq.\eqref{drivingprof}, but employ qualitatively similar driving profiles
\begin{equation}\label{eq:drivingprofilesnc}
\xi_s(t)=\mathcal{E} e^{i\varphi_s}e^{-i\omega_c t}(e^{i 2\omega_m t }+e^{-i 2\omega_m t })\ ,
\end{equation}
with phase shifts $\varphi_s$ whose form is to be determined.\\
Thanks to the chosen detuning, the first order Magnus term ${\cal M}_1$ vanishes irrespectively of the choice for the $\varphi_s$.
The second and third order contribution to the generator of the dynamics over $N$ periods read ${\cal M}_2=\sum_{s=1}^N{\cal M}_2^{(s)}$ and ${\cal M}_3=\sum_{s=1}^N{\cal M}_3^{(s)}$ -- in general, there would be contributions resulting from non-commutativity of ${\cal M}_{1/2}^{(s)}$ and ${\cal M}_1^{(l)}$, but in the present case those do not exist because ${\cal M}_1^{(s)}$ vanishes.

Even though ${\cal M}_2^{(s)}$ and ${\cal M}_3^{(s)}$ display a rather complicated form reflecting the complex dynamics induced by the non-linear Hamiltonian, it is still possible to ensure the desired goals of a product state with an empty cavity and a non-classical state of the mirror.
This is achieved requiring every undesired element in $W_s^\dagger {\cal M}_j W_s$ ($j=2,3$) to be proportional to $\exp(\pm i\varphi_s)$ or $\exp(\pm i2\varphi_s)$, which would suggest to adopt the same set of phase shifts we proposed for the creation of squeezed states, i.e. $\varphi_s=2\pi (s-1)/N$.
Some care, however, is in order since preparing non-classical states relies on the dynamics induced by third order terms in the coupling and thus requires a fairly stronger coupling regime.
This makes an experimental realisation more challenging than the creation of squeezed states which is a second order effect.
On the other hand, the final propagator is enhanced by a factor $\eta^2$, so that strong driving can compensate for the weak interaction.
Yet, in the strong driving regime special care needs to be taken in the perturbative expansion: so far we were only concerned with powers of $k$, but for sufficiently large values of $\eta$, a high power of $\eta$ can make a term relevant despite its high order in $k$. A quantitative analysis of the algebra and the perturbative expansion is provided in the Methods section; here we only outline that the propagator contains terms $\propto k^2\eta^2n_c$ which create neither cavity excitations nor light-matter correlations, but which induce a back-action on the dynamics, rotating the field operators at each period and spoiling the effect of the previously engineered phase shifts.
To counteract this effect that undermines the achievement of a separable state at the end of the $N$ driving periods we should modify the phase shift to
\begin{equation}
\varphi_s=\left(\frac{2\pi}{N}+\frac{4\pi}{3}(k\eta)^2\right)(s-1)\ .
\nonumber
\end{equation}
Making use of all the cancellations, we thus arrive at the desired separable propagator $V(TN,0)= V_c(N)\otimes V_m^{(3)}(N)$ with
\begin{equation}\begin{split}\label{eq:prop3ord}
V_m^{(3)}(N)&=\exp\left(-\frac{\pi}{3} i\ Nk^3\eta^2\ Q_m\right)\ ,\quad \mathrm{and}\\
V_c(N)&= \mathrm{exp}\left(2\pi i \ k^2\left(n_c^2+\frac{2}{3}\eta^2n_c\right)N\right)\  ,
\end{split}\end{equation}
defined in terms of the cubic operator
\begin{equation}\label{qm}
Q_m=\left(X_m+i\frac{P_m}{\sqrt{3}}\right)^3+\left(X_m-i\frac{P_m}{\sqrt{3}}\right)^3+\frac{3}{2}X_m\ .
\end{equation}

In contrast to the well characterised squeezed states discussed above, it is not clearly established what type of states are generated by $Q_m$.
Hence, we construct $V_m^{(3)}(N)$ numerically in a truncated Hilbert space including up to $80\times10^3$ excitations. As prototype for discussion, we consider the state $\ket{\Psi(20)}=V_m^{(3)}(20)\ket{0}$ obtained after $N=20$ periods of driving with the mirror initially in its ground state.
As specific parameter values we choose $\eta=20$ and $k=1/60$ consistently with the perturbative expansion and with up-to-date experimental achievements \cite{kippenberg2008, teufel2011, aspelmeyer2014}).

\section*{Discussion}

\begin{figure*}[t]
\centering
\includegraphics[width=1\textwidth]{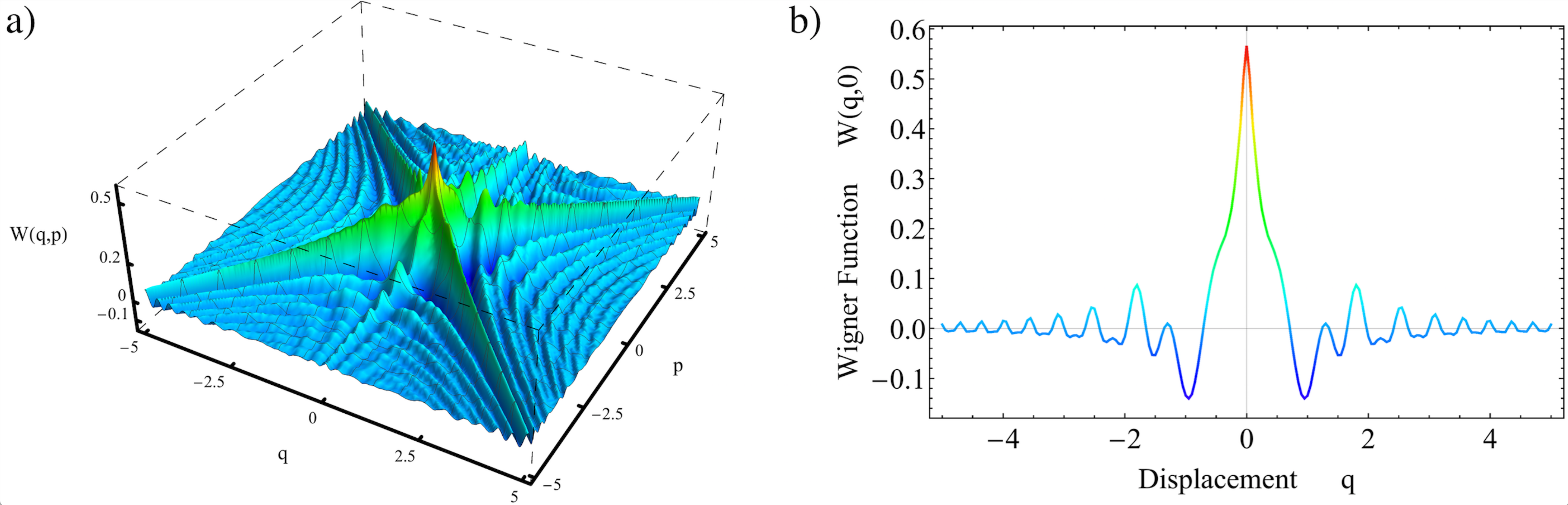}
\caption{a) 3D Wigner function of the mirror after $20$ driving periods and b) its profile when it is cut by the plane $p=0$. The experimental parameters are set as $\eta=20$, $k=1/60$ and the resulting average population is $\langle \bd b\rangle\simeq 20$.}\label{wigner}
\end{figure*}

Since non-linear Hamiltonians tend to generate highly non-classical states, it is instructive to analyse the states that are accessible with the present control scheme in terms of commonly employed measures of non-classicality.
A particularly intuitive approach can be derived in terms of
the Wigner function $W(q,p)=\frac{1}{\pi}\int_{-\infty}^{\infty}\langle q+y|\rho| q-y\rangle e^{-2ipy}dy$, which is a quasi-probability distribution in phase space spanned by momentum and displacement variables $p$ and $q$.
Fig.~\ref{wigner}a) depicts the Wigner function for the state $\ket{\Psi(20)}\bra{\Psi(20)}$.
Quantumness can be characterised by oscillations of $W(q,p)$,
where high-amplitudes of short-wavelength oscillations including negative values imply deep quantum mechanical behavior.
As one can see, the Wigner function of $\ket{\Psi(20)}$ features short wavelength oscillations with large amplitudes.
This is visible on a more quantitative level also in Fig.~\ref{wigner}b) which shows the cut $W(q,0)$ through the Wigner function.

\begin{figure}[b!]
\centering
\includegraphics[width=0.48\textwidth]{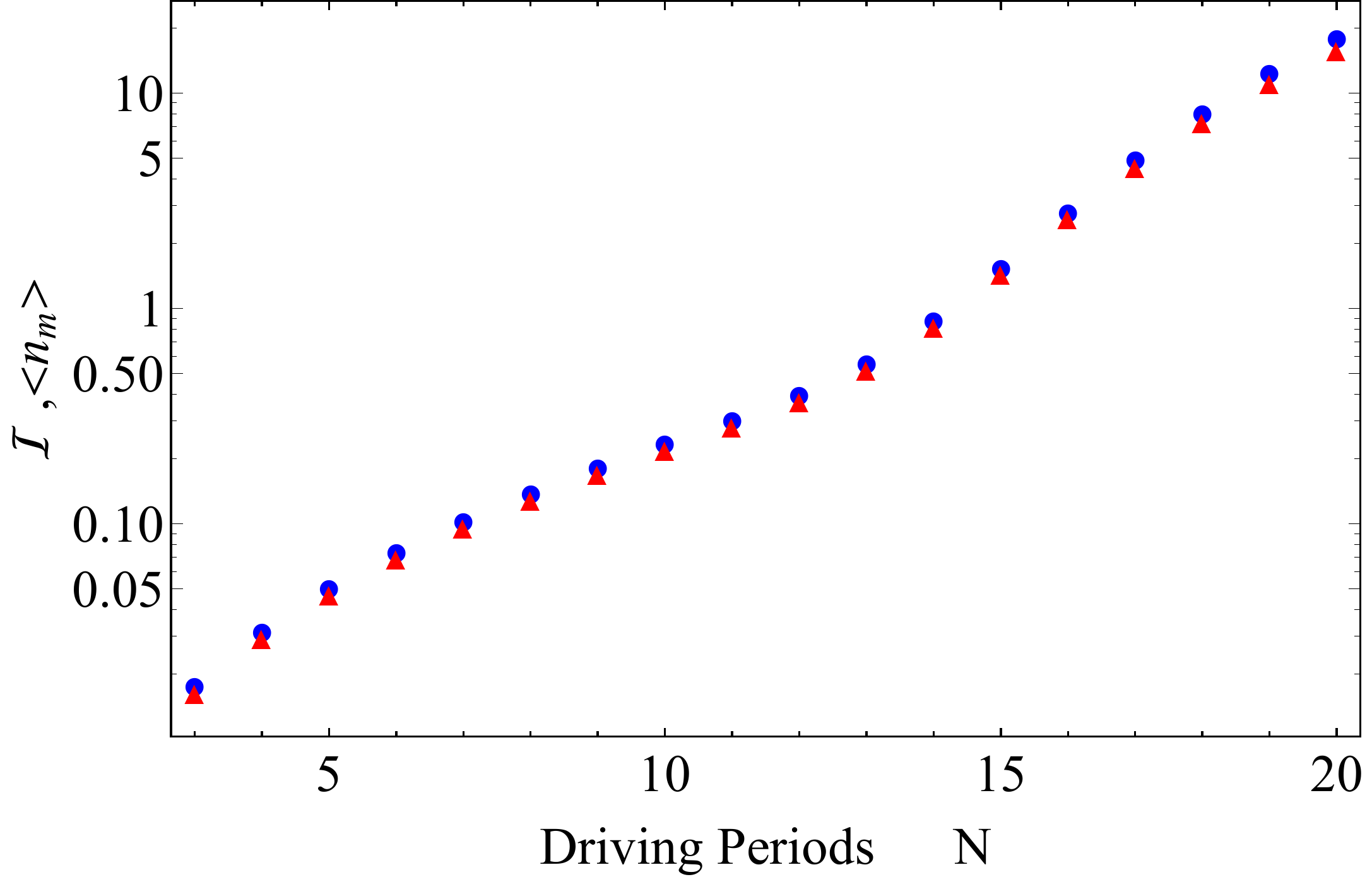}
\caption{Comparative plot of the non-classicality $\mathcal{I}$ (red triangles) and the average number of mechanical excitations $\langle n_m\rangle$ (blue dots) as functions of the number of driving periods. The experimental parameters are set as $\eta=20$, $k=1/60$.}\label{quantumdim}
\end{figure}

In order to provide a quantitative estimate of the quantumness, we resort to the \textit{measure of non-classicality} \cite{lee2011}
\begin{equation}\label{quantumest}
{\cal I}=-\frac{\pi}{2}\int dp\ dq\ W(q,p)\left(\frac{\partial^2}{\partial q^2}+\frac{\partial^2}{\partial p^2}+1\right)W(q,p)\ ,
\end{equation}
that quantifies fast oscillations of the Wigner function $W(q,p)$.
This non-classicality ${\cal I}$ lies in the interval $\mathcal{I}\in[0,\langle n\rangle]$, where $\langle n\rangle$ is the average number of excitations in the system. The minimal value ${\cal I}_{min}=0$ is obtained for classical states like Gaussian or thermal states, while purely quantum states, such as for example Fock and cat states, yield the maximum value of ${\cal I}_{max}=\langle n\rangle$.
We deem ${\cal I}$ a more suitable figure of merit than macroscopicity, $\mu$ \cite{nimmrichter2013} discussed in the introduction, since $\mu$ depends on system parameters like the particle mass, and thus, to a large extent characterises the experimental achievement of a challenging experiment with a massive object. ${\cal I}$ however, reflects solely on the conceptual added value of the present control scheme.

Fig.\ref{quantumdim} depicts $\mathcal{I}$ as a function of the driving time expressed in units of mechanical periods (red triangles) together with the average population $\langle n\rangle$ (blue circles).
Both quantities have an approximately exponential growth, so that highly excited, non-classical states can be prepared very quickly (we obtain $\langle n\rangle\simeq 18$ after $N=20$ periods). Moreover, quantumness nearly saturates the bound ${\cal I}_{max}$ imposed by the population, which witnesses the rapid evolution towards states of macroscopic character as well as their close-to-maximal non-classicality.

So far, we have discussed an idealised situation with unitary dynamics and no experimental imperfections.
An extensive analysis of the resilience against experimental errors can be found in the Supplementary Materials.\cite{supplementary}.
In particular, we analytically show how the driving profiles ensure robustness against optical and mechanical damping, as well as we provide evidence of the small detrimental effect of decoherence by numerically solving the full Lindbladian master equation.
We also demonstrate that there is no fundamental need to require ground state cooling preparation of the mirror, obtaining significative negative values of the Wigner function for an initial state with $\langle n_m^{th}\rangle=1$, \textit{i.e.} substantially above what is already achieved with sideband cooling.
Importantly, also substantial deviations from the step-like phase shifts would not prevent the achievement of a highly non-classical state of the mirror and an empty cavity. Besides, we will argue that the proposed laser driving pattern can be accessed with state of the art technology since phase shifts $\varphi_s$ need to be implemented on long time scales that are of the order of $1/\omega_m$.

Thanks to the resilience to experimental imperfections, the massive mirror could be potentially used as continuous variable quantum memory, as it has already been proposed in Ref.\cite{chan2011}, or as probe for decoherence \cite{romero2011, bahrami2014, abdi2016}.
The non-classicality $\mathcal{I}$ of the mirror is an extremely sensitive indicator of any type of mechanical decoherence and is thus ideally suited to probe fundamental physics such as gravitationally induced effects on the mechanical motion or continuous spontaneous localization \cite{ghirardi1990}.

It should also be highlighted that the utilized approach to find optimal driving patterns can be easily extended to higher orders in the Magnus expansion and correspondingly longer propagation times and/or larger coupling $k$. There is indeed no theoretical restriction to an adaptive fine tuning of the laser profiles to cancel undesired coupling terms in the evolution. This would give rise to more highly excited states and hence to measurable quantum effects also in case of higher initial thermal noise, pushing the initial cooling condition beyond the requirement $\langle n_m^{th}\rangle \lesssim 1$.
The present control scheme is also not necessarily restricted to the mirror-cavity setup discussed here, but similar techniques are suitable for a variety of systems that share similar non-linear hamiltonians such as atomic spin ensembles, trapped atoms or levitated nanoparticles \cite{myatt2000, julsgaard2001, scala2013, hammerer2009}.

\section*{Methods}

We provide full details on the reconstruction of the propagator induced by the Hamiltonian in Eq.~(1) together with the driving profile in Eq.~(5).
Let us start by recalling the Magnus expansion for the propagator $V(T,0)=\exp(-i\sum_j{\cal M}_j^{(1)})$ over the first mechanical period. Thanks to the chosen detuning, the first order Magnus term ${\cal M}_1$ vanishes irrespectively of the choice for the $\varphi_s$.

The second and third order terms are
\begin{equation}
\begin{split}\label{suppmagnussuperp}
{\cal M}_2^{(1)}&=\pi k^2\left(m_2^{c}+m_2^{I}-\frac{29}{60}\eta^4\right)\ ,\hspace{.3cm}\mbox{with}\\
m_2^{c}&=-2 {n}_c^2+\frac{1}{3}\eta^2(X_c^2-6 {n}_c)\hspace{.3cm}\mbox{and}\\
m_2^{I}&=\eta P_c(b^2+{\bd}^2)\ ;
\end{split}
\end{equation}
as well as
${\cal M}_3^{(1)}=\frac{\pi}{3}k^3\eta\left(m_3^{m}+m_3^{I}\right)$, with
\begin{equation}
\begin{split}
\label{suppmagnussuperp3}
m_3^{I}=&\left[14i(\ad n_c -n_c a)-\left(\frac{36}{5}\eta^2+4\right)P_c\right]X_m\\
&+\left[3X_c+6i\eta(a^2-{\ad}^2)\right]P_m-\frac{3}{4} P_c Q_m\ ,
\end{split}
\end{equation}
and $m_3^{m}=\eta\  Q_m$, with $Q_m$ defined in Eq.\eqref{qm}).\\
Exploiting the composition property we write the identity $V(TN,0)=\prod_{s=1}^N W_s^\dagger V(T,0)W_s=\prod_{s=1}^N\exp(-i{\cal M}^{(s)})$ with ${\cal M}^{(s)}=W_s^\dagger {\cal M}^{(1)}W_s$ and $W_s=\exp(-in_c\varphi_s)$. Choosing the same set of phase shifts as for the creation of mechanical squeezed states, \textit{i.e.} $\varphi_s=2\pi/N(s-1)$, one obtains
\begin{equation}\begin{split}\label{supppropagator}
V(TN,0)&= V_c(N)\otimes V_m^{(3)}(N),\quad \mathrm{with}\\
V_c(N)&=\mathrm{exp}\left(2\pi i \ k^2\left(n_c^2+\frac{2}{3}\eta^2n_c\right)N\right)\ ,\quad \mathrm{and}\\
V_m^{(3)}(N)&=\exp\left(-\frac{\pi}{3} i\ N k^3\eta^2\ Q_m\right)\ .
\end{split}\end{equation}
Since the propagator for the mirror $V_m^{(3)}$ scales as $k^3\eta^2$, the cubic dependence on $k$ will make an experimental realisation more challenging than the creation of squeezed states which is a second order effect.
Given the quadratic dependence on $\eta^2$, however, strong driving can compensate for the weak interaction.
Yet, in the strong driving regime special care needs to be taken in the perturbative expansion since a high power of $\eta$ can make a term relevant despite its high order in $k$.
In particular, ${\cal M}_2^{(1)}$ in Eq.\eqref{suppmagnussuperp} contains terms $\sim(k\eta)^2n_c$, and terms $\sim(k\eta)^4$ resulting from the commutators $[{\cal M}_2^{(s)},{\cal M}_2^{(l)}]$.
This is not directly a severe issue for the state preparation, since such terms describe neither an interaction between cavity and mirror nor non-interacting dynamics of the mirror. They do induce, however, a perturbative rotation of the cavity field.
As a result of that, the propagator at the end of the driving time does not factorize into individual propagators of mirror and cavity. It thus becomes necessary to change the driving profile accordingly to compensate for this effect.\\
In order to do so, it is instructive to rewrite the propagator over the first interval, neglecting terms of order $k^4\eta^j$ with $j<4$ and terms of order $k^j$ with $j>4$, as
\begin{equation}
\begin{split}\label{newphases}
V(T,0)&\simeq e^{i\frac{4}{3}\pi k^2\eta^2n_c}e^{-i\mathcal{\tilde{M}}^{(1)}}\ , \ \mathrm{with}\\
\tilde{\mathcal{M}}^{(1)}&=\pi k^2(\tilde{m}_2^c+m_2^I)+\mathcal{M}_3^{(1)}\ , \ \mathrm{and}\\
\tilde{m}_2^c&=-2n_c^2+\eta^2({\ad}^2+a^2+1)/3\ ,\\
m_2^{I}&=\eta P_c(b^2+{\bd}^2)\ ,
\end{split}
\end{equation}
where the term $\exp(i\frac{4}{3}\pi k^2\eta^2n_c)$ in the expression for $V(T,0)$ -- and similarly for $V(sT,(s-1)T)$ -- is the undesired rotation.
The propagator over $N$ periods can be written as
\begin{equation}
V(TN,0)=W_{N+1}^\dagger \left(\prod_{s=1}^{N} W_{s+1}W_s^\dagger V(T,0)\right)W_1\ ,
\nonumber
\end{equation}
and we should choose the $W_s$ such that the prefactors $W_{s+1}W_s^\dagger$ cancel the term $e^{i\frac{4}{3}\pi k^2\eta^2n_c}$ in Eq.~\eqref{newphases}.
This is achieved with the set of phases
\begin{equation}
\varphi_s=\left(\frac{2\pi}{N}+\frac{4\pi}{3}(k \eta)^2\right)(s-1)\ ,
\nonumber
\end{equation}
which counterbalances exactly the phase shift $\Delta=\frac{4\pi}{3} k^2 \eta^2$ that the cavity experiences through the driving over each period as described in Eq.\eqref{newphases}.
With this, the propagator reads
\begin{equation}
V(TN,0)=\prod_{s=1}^N \exp\left(e^{\frac{2\pi i}{N}(s-1)n_c}\tilde{\cal M}^{(1)}e^{-\frac{2\pi i}{N}(s-1)n_c}\right)\ ,
\end{equation}
and the basic principles discussed in main text for the cancellation of all the interaction and cavity excitation terms apply.
Quite importantly, however, the terms $\sim(k\eta)^2n_c$ no longer appear, and the only remaining contribution scaling as $\sim(k\eta)^2$ in $m_2^c$ in Eq.~\eqref{newphases} is the polynomial ${\ad}^2+a^2+1$.
Operators ${\ad}^2$ and $a^2$ cancel out exactly in the summation over the $N$ periods thanks to the specific set of phase shifts, and the `$+1$'  brings an irrelevant global phase.
The terms $\sim(k\eta)^4$ arising from the commutators $[{\cal M}_2^{(s)},{\cal M}_2^{(l)}]$ (which are of the form $[a^2,{\ad}^2]=4n_c+2$) contribute either to the global phase or to a global final rotation in $V_c$.
Lastly, we will see that the only term $\sim(k\eta)^4$ in ${\cal M}_4^{(1)}$ depends on $P_c$ and averages out in the summation over the $N$ periods.
The explicit final form of Eq.\eqref{newphases} coincides with Eq.\eqref{supppropagator} and the one given in the main text.

\subsection*{Data availability}
This is a theoretical paper and there is no experimental data available beyond the numerical simulation data described in the paper.

\section*{Acknowledgements}
We are grateful for fruitful discussions with Michael Vanner and Benjamin Dive.
This work was supported by the European Research Council within the project odycquent.

\section*{Author contributions}
All authors made substantial contributions and were involved in drafting and writing up the work.

\section*{Additional Information}

\textbf{Supplementary Information} accompanies the paper with an extensive discussion on the resilience of the state preparation to optical and mechanical decoherence, initial mechanical thermal noise and imperfect optical driving.

\textbf{Competing interests} The Authors declare that they do not have any competing interest.

\begin{widetext}
\clearpage

\begin{center}

\textbf{\large Supplementary material for:\\
\vspace{0.2 cm}
Deterministic preparation of highly non-classical macroscopic quantum states}\\

\vspace{0.4 cm}

Ludovico Latmiral and Florian Mintert\\
\textit{QOLS, Blackett Laboratory, Imperial College London, London SW7 2AZ, United Kingdom}

\end{center}

In this supplementary material, we provide an extensive discussion on the resilience of the state preparation to optical and mechanical decoherence, initial mechanical thermal noise and imperfect optical driving.

\vspace{0.8 cm}

\end{widetext}

\setcounter{equation}{0}
\setcounter{figure}{0}
\setcounter{table}{0}
\setcounter{page}{1}
\makeatletter
\renewcommand{\theequation}{S\arabic{equation}}
\renewcommand{\thefigure}{S\arabic{figure}}
\renewcommand{\bibnumfmt}[1]{[S#1]}
\renewcommand{\citenumfont}[1]{S#1}

\subsection{Perturbative regime}

\begin{figure}[b!]
\centering
\includegraphics[width=0.48\textwidth]{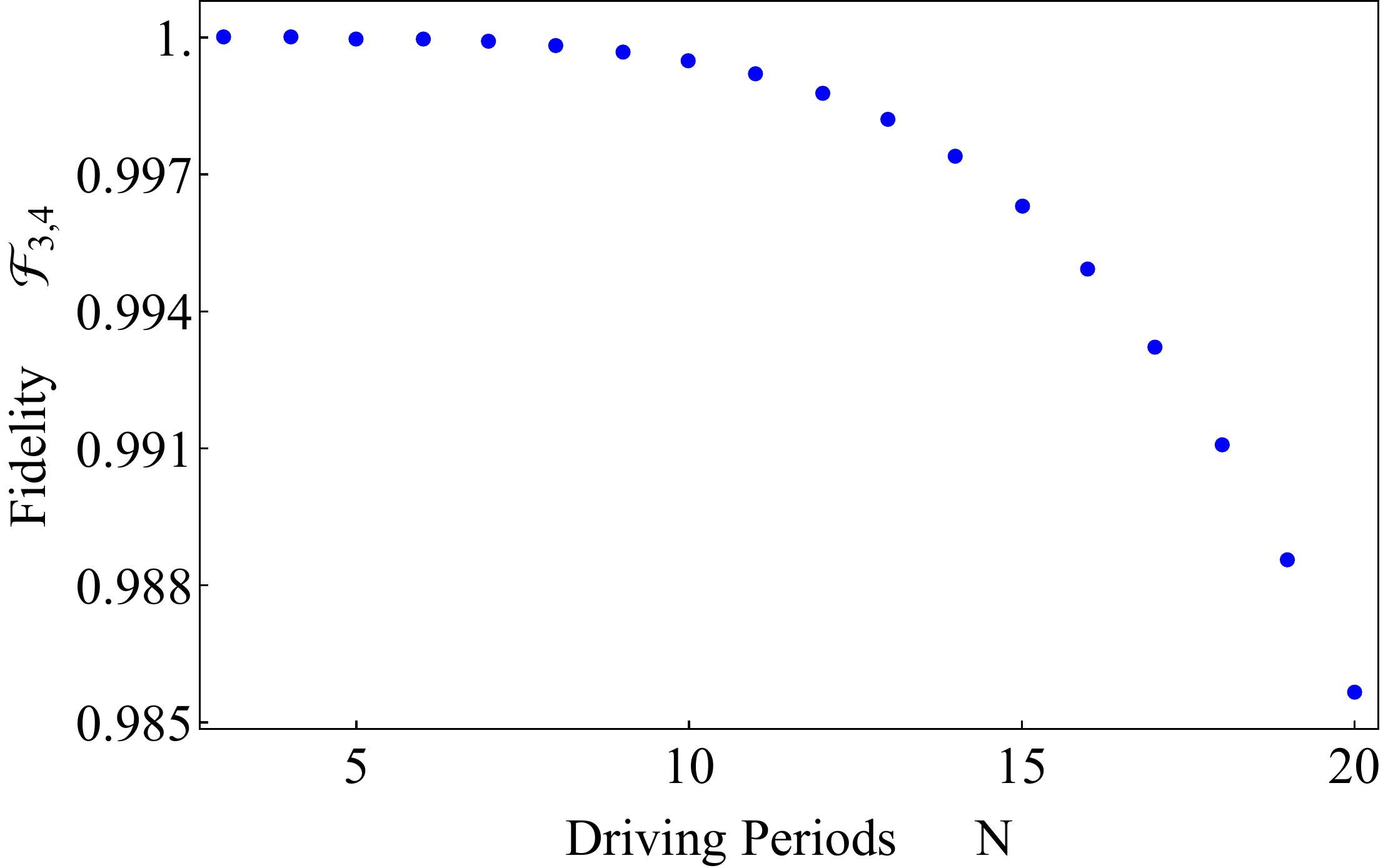}
\caption{Fidelity between the state of the mirror computed via a third and a fourth order Magnus expansion as a function of the integration time, expressed in terms of mechanical driving periods. The experimental parameters are set as $\eta=20$, $k=1/60$. The graph indicates that high fidelity is obtained up to $20$ driving periods: $\mathcal{F}_{3,4}\gtrsim 0.985$.}
\label{fidelity}
\end{figure}

The Magnus expansion relies on the perturbative regime $k=\frac{g}{\omega_m}\ll 1$, and it is essential to gauge the range of applicability of the perturbative approximation. To this end, we compare the dynamics obtained with the Magnus expansion in third- and fourth-order approximation.\\
Applying the same procedure we have presented so far, we find the dynamics of the mirror after $N$ mechanical periods at fourth order
\begin{equation}\label{cosuperpositionS}\begin{split}
V_m^{(4)}(N)=&e^{-\frac{\pi}{3} i\ N k^3\eta^2 Q_m}\ e^{-i\frac{\zeta^{(4)}}{2}({\bd}^2+b^2)^2}\\
\times &e^{i\pi k^4 \eta^2\ N \left[\frac{124\eta-5}{20}(b^2+{\bd}^2)+\frac{575+634\eta}{90}\bd b\right]}\ ,
\end{split}\end{equation}
with $\zeta^{(4)}=(\pi k^2\eta)^2N\cot(\pi/N)$.
Hence, we recall an important figure of merit, \textit{i.e.} the state fidelity between two quantum states $\varrho_A$ and $\varrho_B$, which is defined as
\begin{equation}\label{defstatefidelity}
\mathcal{F}_{A,B}=\left(\Tr\sqrt{\varrho_A^{1/2}\varrho_B\varrho_A^{1/2}}\right)^2\ .
\end{equation}
This quantity can be used to provide an estimate of the accuracy of the state preparation computed at the third order in Magnus, $\varrho_3$, by comparing it with $\varrho_4$, obtained by numerically propagating Eq.\eqref{cosuperpositionS}.
Fig.~\ref{fidelity} depicts $\mathcal{F}_{3,4}$ as function of the driving time in terms of mechanical periods. We infer that the deviations between $\varrho_3$ and $\varrho_4$ are in the permille regime for the first $10$ driving periods, and even at $N=20$, the third order approximation is accurate within $\simeq 1\%$.
We deem an error of $1\%$ below the accuracy of what could be achieved experimentally within the next years, and thus feel that the perturbative treatment is highly adequate for the present purpose.

\subsection{Experimental imperfections}\label{experimentalimperfections}

It is appropriate to gauge how unavoidable experimental imperfections will affect the desired process.
To this end, we analyse the impact of optical and mechanical decoherence, initial thermal excitations in the mirror and imperfect phase shifts of the driving fields.
Since the dynamics takes place in a high-dimensional Hilbert spaces of both mechanical and optical degree of freedom, this discussion is necessarily restricted to approximate methods.
Together with the verification of the quality of the perturbative expansion (discussed in \ref{sec:opticallosses} below), this seems adequate to estimate the order of magnitude of imperfections.

\subsubsection{Optical decoherence}
\label{sec:opticallosses}

Due to experimental difficulties, the most delicate aspect affecting the unitarity of the evolution, and consequently the preparation of the desired mechanical state, is attributable to optical losses. The leakage of photons from the cavity is quantified by the decay rate $\kappa$, which is defined as the inverse of the time light remains in the cavity. 
A common approach to include the effects of such photon losses in the dynamics is to express the evolution of the system density matrix $\rho$ in terms of the Master equation $\dot{\rho}=-i\left[H,\rho\right]+\kappa\mathcal{L}_a[\rho]$, where $H$ is the system Hamiltonian and $\mathcal{L}_a[\rho]$ is the Lindblad operator $\mathcal{L}_a[\rho]=(a\rho\ad -\{\ad a,\rho\}/2)$.\\
When operating in the so called \textit{resolved sideband} regime with $\kappa \ll \omega_m$, that is realised in many current experiments \cite{thompson2008, teufel2011, aspelmeyer2014, wollman2015}, the impact of photon loss can be captured by looking at the perturbative solution of the Master equation. To this end, we consider the Master equation for $\tilde{\rho}=V^\dag(t)\rho V(t)$, where $V(t)$ satisfies $i\dot{V}=HV$, as obtained in the main text (see Eq.$6$).

Solving this perturbatively, yields the contribution of photon loss to the dynamics in terms of powers of $\kappa/\omega_m$. The integration for a generic time $t$ assumes a rather complex form and correlations between field and mirror are created because of dissipation. However, thanks to the very specific bi-chromatic driving pattern many terms cancel out at the end of the evolution. Actually, our proposed set of constant phase shifts $\{\varphi_k\}$ is crucial to suppress the majority of these unwanted non-unitary and de-coherent contributions, including all correlation terms proportional to the driving amplitude $\eta$.

More specifically, in leading order in $k$ and $\kappa$, we obtain at $t=\mathrm{NT}$ an expression that is completely independent of $\eta$ and is thus well suited to describe the strong driving regime
\begin{equation}
\label{leak}
\tilde{\rho}(NT)\simeq\tilde{\rho}(0)+\kappa NT \left(\tilde a\tilde{\rho}\tilde a^\dagger-\frac{1}{2}\{\tilde{a}^\dagger \tilde{a},\tilde{\rho}\}\right)\ ,
\end{equation}
with
\begin{equation}\label{missdisp}
\tilde a =ae^{\frac{g}{\omega_m}(b - \bd)}\ .
\end{equation}
This result has a very clear physical interpretation: since the light-matter interaction conditionally displaces the mirror by an amount proportional to the number of photons in the cavity, each photon that has leaked out of the resonator should then be matched with a \textit{missing} displacement $e^{g(b-\bd)/\omega_m}$ of the mirror as indicated in Eq.\eqref{missdisp}.

\begin{figure}[b!]
\centering
\includegraphics[width=0.48\textwidth]{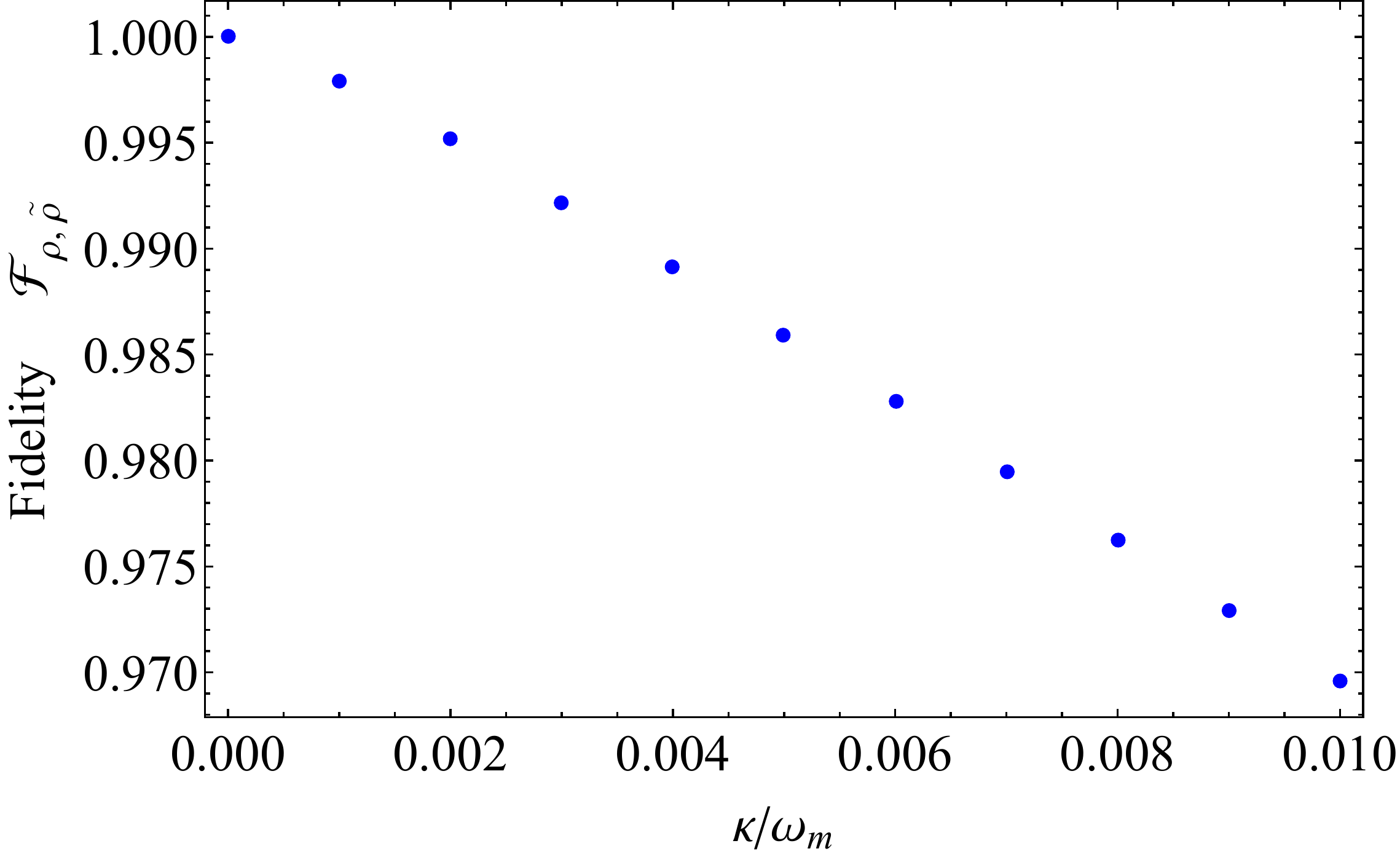}
\caption{Fidelity between the final state of the system (cavity plus mirror) in case of photon losses and the ideal scenario as a function of the cavity decay rate $\kappa$ for an evolution lasting $20$ mechanical periods.}\label{leaking}
\end{figure}

Fig.\ref{leaking} depicts the state fidelity (see Eq.\eqref{defstatefidelity}) after $N=20$ mechanical periods as a function of the ratio $\kappa/\omega_m$ between the full state of the system (cavity plus mirror) obtained in the leaking scenario in Eq.\eqref{leak} and the ideal one discussed in the main text.
As one can see, a loss rate satisfying $\kappa/\omega_m < 10^{-2}$ results in a reduction of the fidelity by $\lesssim 3\%$.
This condition, together with the strong driving regime, is in accordance with Ref.\cite{miao2009}, where the resolved sideband regime and the condition $g/\kappa>1$ were theoretically derived as requirements to resolve the \textit{granularity} of the photon stream and fully exploit the non-linearity of the system to observe purely quantum features.

\subsubsection{Thermal initial state of the mirror}

Since the evolution operator $V$ (see Eq.$6$ of the main text) factorises into a propagator for the mirror and a propagator for the cavity, one obtains a product state of mirror and cavity for any initial product state.
That is, there is no fundamental need to require the mirror to be initially cooled exactly to the ground state, but initial thermal excitation of the mirror will affect the non-classicality of the final state.

\begin{figure}[h!]
\centering
\includegraphics[width=0.48\textwidth]{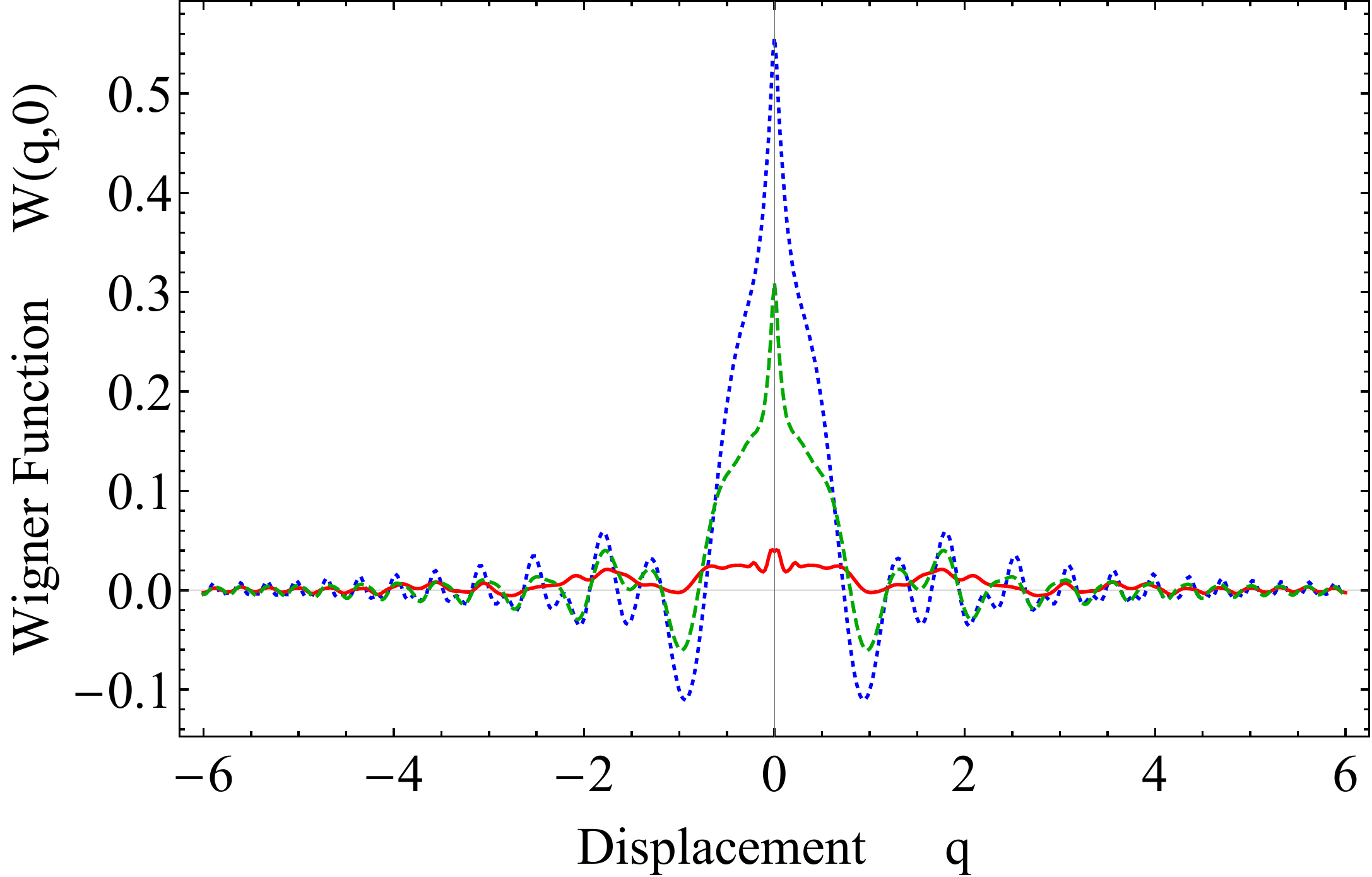}
\caption{Comparison of the cut profiles $p=0$ of the Wigner function of the state of the mirror after an evolution lasting $20$ mechanical periods with the mirror initially in its ground state (blue dotted line) and two thermal states with respectively $\langle n_m^{th}\rangle \sim 1$ (green dashed line) and $\langle n_m^{th}\rangle \sim 10$ (red line). The experimental parameters are set as $\eta=20$, $k=1/60$ and $\omega_m=2\pi\times 10^7 \mathrm{Hz}$.}\label{wignerthermal}
\end{figure}

Fig.\ref{wignerthermal} depicts cuts through the Wigner function of the mirror after $20$ periods of driving for different initial thermal populations with $\langle n_m^{th}\rangle=1$ and $\langle n_m^{th}\rangle=10$,
{\it i.e.} above the experimental threshold of $\langle n_m^{th}\rangle \sim 0.2$ achievable with sideband cooling (at a mechanical frequency $\omega_m=2\pi\times 10^7 \mathrm{Hz}$) \cite{teufel2011,clark2017}.
The strong oscillatory behaviour with negative values of $W$ is clearly displayed for an initial state with $\langle n_m^{th}\rangle=1$.
Only for $\langle n_m^{th}\rangle=10$, {\it i.e.} substantially above the limits of side-band cooling, the quantum mechanical features are mostly outshined by the thermal contributions.

\subsubsection{Mechanical decoherence}\label{mechanicaldecoherence}

The main source of mechanical decoherence for a cooled optomechanical resonator  arises from mechanical damping, which is characterised by the rate $\gamma_m$ at which a phonon excitation is lost in the environment. In case of non-zero temperature, however, the unwanted absorption of thermal excitations should also be taken into account. 
The process is conceptually analogous to optical photon losses from the cavity which happen at rate $\kappa$.
Current experiments have achieved mechanical damping substantially below photon loss ($\gamma_m\ll \kappa$), what suggests that mechanical decoherence will not be a limiting factor.
Since, however, highly non-classical, coherent superpositions of macroscopically distinct states are particularly sensitive to decoherence, a critical assessment of motional decoherence is in order.
To this end we analyse the dynamics induced by the Master equation $\dot{\rho}=-i[H,\rho]+\gamma_m [(\langle n_m^{th}\rangle+1) \mathcal{L}_b[\rho]+\langle n_m^{th}\rangle \mathcal{L}_{\bd}[\rho]]$, where $\mathcal{L}_b$ and $\mathcal{L}_{\bd}$ are the Lindblad operators for phonon absorption and emission, defined similarly to Sec.\ref{sec:opticallosses}.
Thanks to high mechanical quality factors $Q=\omega_m/\gamma_m \gg 1$ being achieved in various experimental realizations, a perturbative solution of the Master equation provided reliable estimates.
At the end of the state preparation, the system state $\tilde{\rho}$ in the frame defined by $V(t)$ (defined in Sec.\ref{sec:opticallosses}) at leading order in $\gamma_m$ and $\kappa$ reads
\begin{equation}
\tilde{\rho}(NT)=\tilde{\rho}(0)+\gamma_m \left[(\langle n_m^{th}\rangle+1) \mathcal{L}_m[\tilde{b},\rho]+\langle n_m^{th}\rangle \mathcal{L}_m[\tilde{\bd},\rho]\right]\ ,
\end{equation}
with
\begin{equation}
\tilde{b}=b-\kappa(\ad a + \frac{\eta^2}{2})\ .
\end{equation}
\begin{figure}[b!]
\centering
\includegraphics[width=0.48\textwidth]{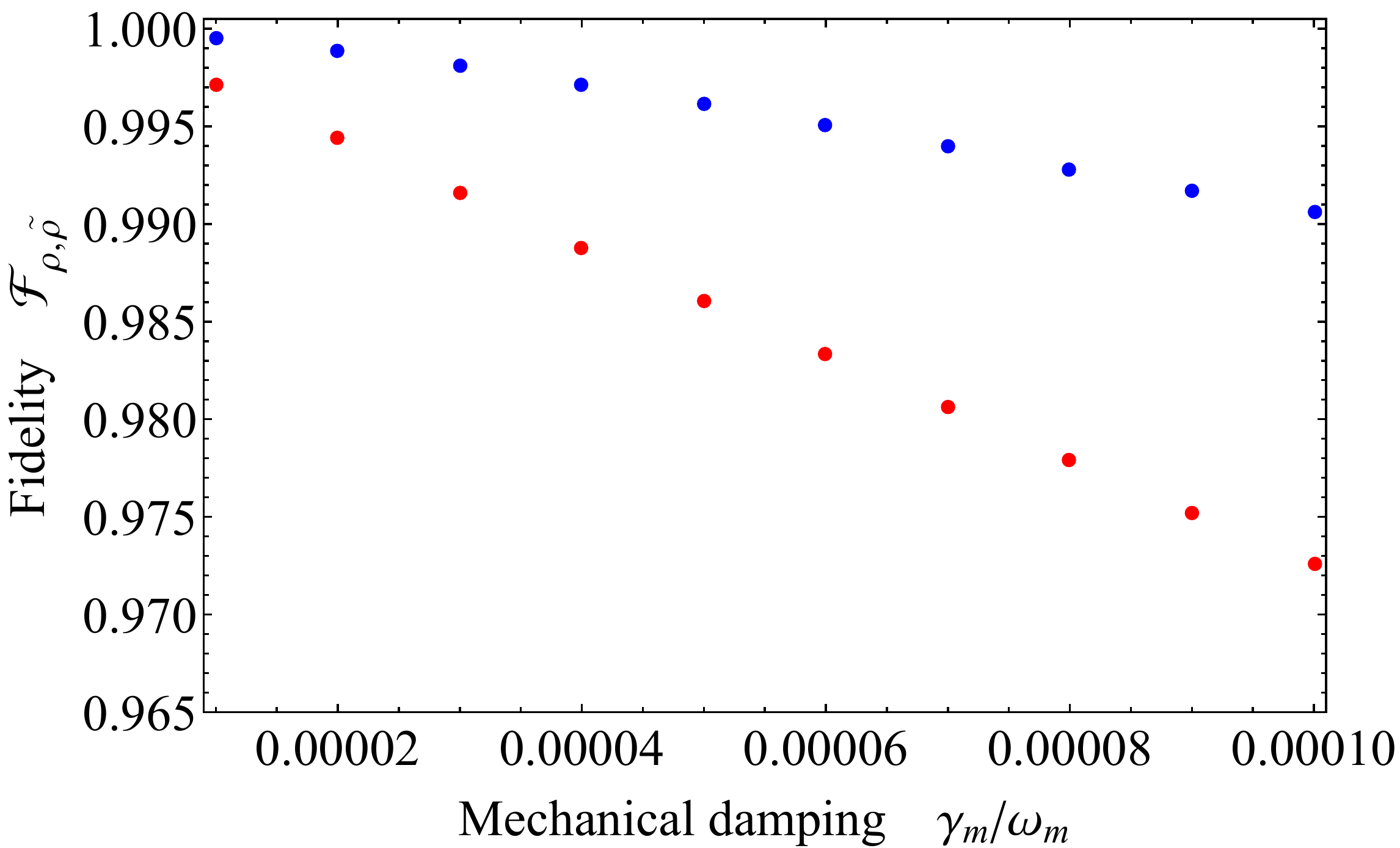}
\caption{Fidelity between the final state of the system (cavity plus mirror) in case of a mechanical damped evolution and the ideal scenario as a function of the mechanical damping $\gamma_m/\omega_m$. Blue points are referred to the zero bath temperature case, while red points to $\langle n_m^{th}\rangle\sim 1$. The total evolution is supposed to last $20$ mechanical periods with the mirror initially in its ground state and the dimensionless driving and coupling respectively set as $\eta=20$ and $k=1/60$.}\label{mechanicaldamp}
\end{figure}
The result is compact, as satisfactorily, many terms get simplified at the end of the evolution because of the very specific choice of the driving profiles. Fig.\ref{mechanicaldamp} depicts the state fidelity after $N=20$ periods of driving as a function of $Q^{-1}$, both in the case of a zero-temperature environment (in blue) and for a thermal state with $\langle n_m^{th}\rangle\sim 1$ (red), as obtained in recent experiments \cite{teufel2011,clark2017}.
In both cases the impact of mechanical damping on the state fidelity is smaller than in the case of optical decoherence. In particular at zero temperature the impact is almost negligible (within the per-mille regime) and even with thermal noise, reductions of the fidelity are limited to a few percent.

\subsubsection{Optical and mechanical decoherence}

After having analyzed mechanical and optical decoherence separately, we finally assess their combined impact on state preparation in a non-perturbative analysis.
To this end, we numerically solve the full master equation for mirror and cavity for the whole driving time interval, which is necessarily limited to a finite (low) dimensional subspace.
We performed our numerical simulations truncating the cavity field and the mirror respectively to 15 and 35 excitations. This lower truncation of the Hilbert space restricts the range of safely explorable mechanical states to the subset $\langle n_m \rangle \lesssim 5$, which, leaving all others parameters unchanged, requires to account for smaller couplings (we have chosen $k=1/90$).

\begin{figure}[h!]
\centering
\includegraphics[width=0.48\textwidth]{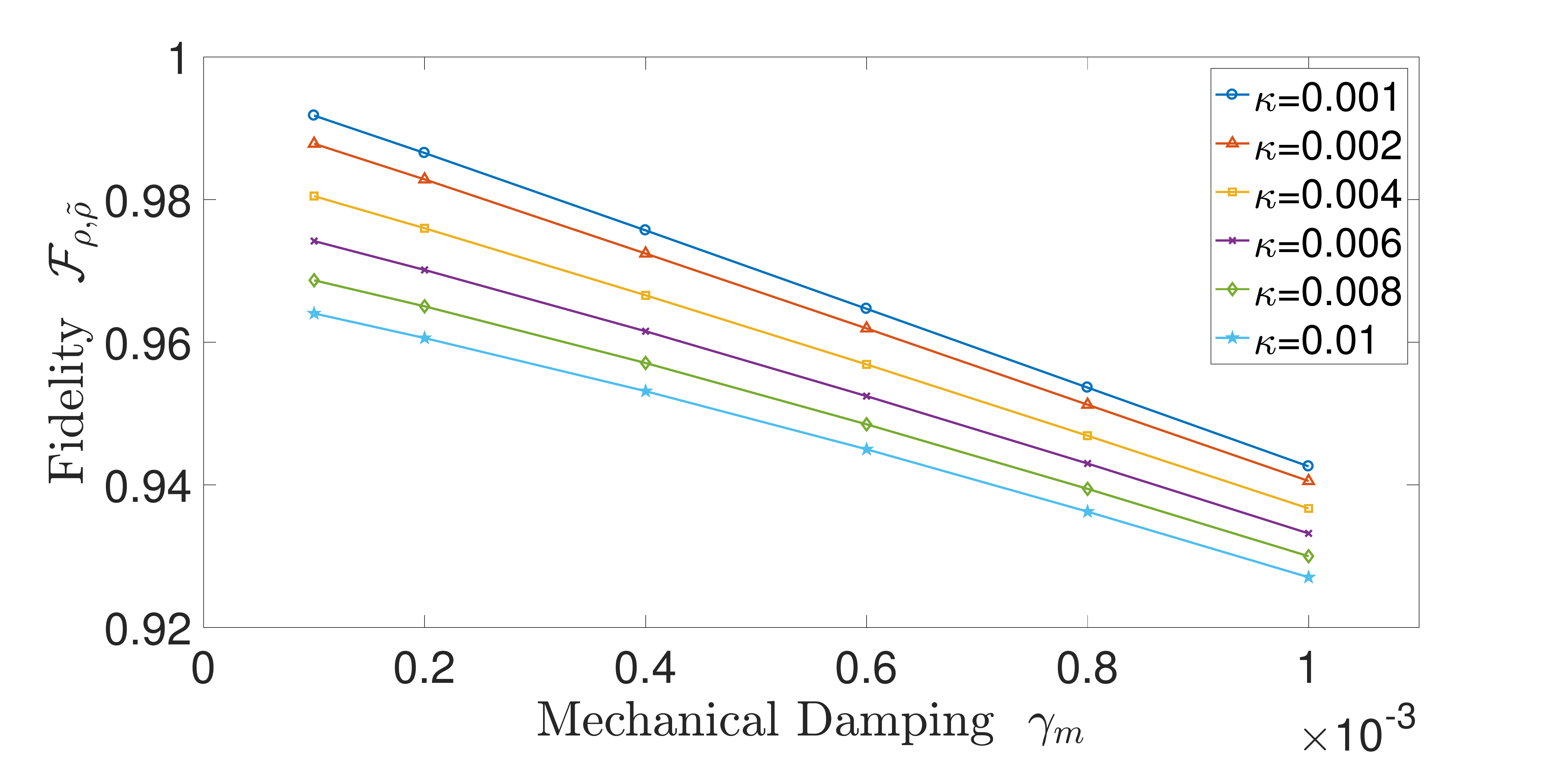}
\caption{Fidelity between the final state of the system (cavity plus mirror), obtained from a numerical simulation of the full master equation, as a function of the mechanical damping $\gamma_m/\omega_m$ and the optical decay rate $\kappa$. We assumed a cooled mechanical oscillator with a thermal bath at $\langle n_m^{th}\rangle\sim 1$. The total evolution is supposed to last $20$ mechanical periods, the dimensionless driving and coupling are set to $\eta=20$ and $k=1/90$.}\label{masterequation}
\end{figure}

We summarize in Fig.\ref{masterequation} the fidelity of the states obtained with a numerical solution of the full master equation accounting for photon losses and phonons absorption and dissipation. Results are in line with analytical discussions provided in Secs. \ref{sec:opticallosses} and \ref{mechanicaldecoherence}. The overall reduction of fidelity remains in the percent regime, consistent with the above perturbative results.

\subsubsection{Laser driving}

\begin{figure}[b!]
\centering
\includegraphics[width=0.48\textwidth]{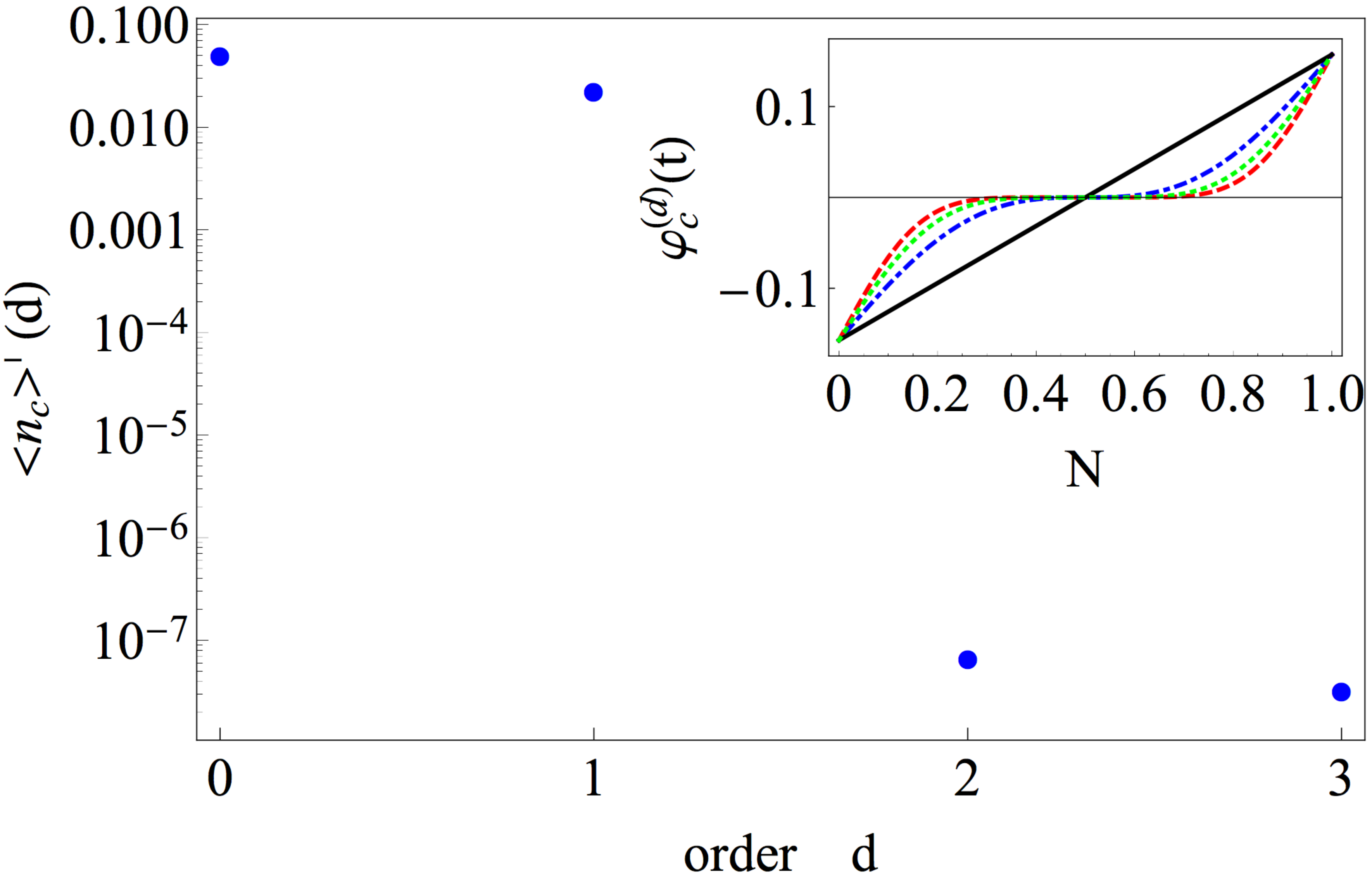}
\caption{Cavity occupation renormalized with respect to the population of the mirror $\langle n_c\rangle^\prime=\langle n_c\rangle/\langle n_m\rangle$ as a function of the order of the decomposition of the step function $d$. In the top-right corner we plot an enlargement of the driving profiles defined by $\varphi_c^{(d)}(t)$ over the first mechanical period: linear approximation with $d=0$ (black line), $d=1$ (blue dashed-dotted), $d=2$ (green dotted) and $d=3$ (red dashed). The experimental parameters are set as $\eta=20$, $k=1/60$, $N=20$.}\label{stepfunction}
\end{figure}

The central goal of deterministic state preparation is achieved through the application of appropriate phase shifts after each period of driving.
Their experimental implementation seems feasible, since the shifts need to be applied on a time-scale that is short as compared to the inverse mechanical frequency. With $\omega_m\sim 10^{6}s^{-1}$, this is orders of magnitude smaller than the optical characteristic frequency.
Accurate control of optical phase has already been demonstrated in Ref.\cite{thom2013}, where phase shift resolutions smaller than $\Delta\varphi=10\mathrm{mrad}$ were achieved.
The technique is based on high-speed fiber optical switchers with switching rate shorter than $1 \mathrm{ns}$, which are already commercially available \cite{thorlabs}.\\
Still, despite the relative simplicity of such scheme, we deem it useful to consider the impact of deviations from the ideal driving profiles with the step-like phase shifts $\varphi_s=\frac{2\pi}{N}(s-1)$. To this end, let us replace the discontinuously evolving phase $\varphi(t)=\frac{2\pi}{N}\sum_s\Theta(t-sT)$ with the continuous function
\begin{equation}
\varphi_c^{(d)}(t)=\frac{2\pi}{N}\frac{t}{T}+\sum_{l=1}^d\varphi_l(t)\ ,
\nonumber
\end{equation}
where $\frac{2\pi}{N}\frac{t}{T}$ is a linearly increasing phase factor and each term $\varphi_l(t)= A_l \sin(l\omega_m t)$ oscillates with frequency $l\omega_m$ and amplitude $A_l$. The set of amplitudes is chosen such that at any order $d$, $\varphi_c^{(d)}(t)$ is tangent to the step function in the centre of the step, \textit{i.e.} for $t=(2j\pi+1)/\omega_m$ with $j\in [0,N-1]$ (see Fig.\ref{stepfunction} for a graphical representation).

Thanks to the continuous time dependance, it is then possible to analytically compute the generator with a Magnus expansion over the entire time window $t=NT$, and subsequently numerically integrate the dynamics over $N$ mechanical periods. Interestingly, we obtain a separable propagator at every order $d$, without correlations between mirror and cavity, and which will still result in deterministic state preparation.

Most importantly, while resorting to the sole linear function $\frac{2\pi}{N}\frac{t}{T}$ single-particle terms of the cavity do not completely cancel out, resulting in a final average population $\langle n_c\rangle \sim 0.2\langle n_m\rangle$, these contributions are efficiently suppressed already at the order $d=3$, when $\langle n_c\rangle \sim O(10^{-7}) \langle n_m\rangle$ (see Fig.\ref{stepfunction}). This is an essential requirement since cavity excitations could potentially prevent the final readout through \textit{back-action-evading} interaction.

Remarkably, the final non-classical mechanical state of the mirror obtained with these imperfect driving pattern presents a very high fidelity ${\cal F}\simeq 0.98$ with the ideal step-like case.

\subsection{Readout}\label{readout}

The final readout of the mechanical motion is a matter that has already been widely analysed theoretically \cite{zhang2003,vanner2015} and implemented experimentally \cite{lei2016,clark2017} with high precision.
The most promising technique to perform quantum state reconstruction is called \textit{back-action-evading} interaction and is based on state transfer. When the mirror is in the state of interest and the cavity is empty, a red detuned laser with frequency $\omega_d=\omega_c-\omega_m$ induces exchange of excitations from the former to the latter.
Hence, tomography of the prepared mechanical state of the mirror can be carried out through homodyne measurement of the light leaking out of the cavity \cite{smithey1993, dariano1994}.\\
Since we ensure that there are no residual correlations between cavity and mirror when the measurement protocol is applied, the desired mechanical quantum state is deterministically read out.

\end{document}